# Scaling Law for Criticality Conditions in Heterogeneous Energetic Materials under Shock Loading

A. Nassar[*], N. K. Rai[*], O. Sen[*], and H.S. Udaykumar[*]

[*]Department of Mechanical Engineering
University of Iowa, Iowa City, IA, USA

**Abstract.** Initiation in heterogenous energetic material (HEM) subjected to shock loading occurs due to the formation of hot spots. The criticality of the hot spots governs the initiation and sensitivity of HEMs. In porous energetic materials, collapse of pores under impact leads to the formation of hot spots. Depending on the size and strength of the hot spots chemical reaction can initiate. The criticality of the hot spots is dependent on the imposed shock load, void morphology and the type of energetic material. This work evaluates the relative importance of material constitutive and reactive properties on the criticality condition of spots. Using a scaling-based approach, the criticality criterion for cylindrical voids as a function of shock pressure, $P_s$ and void diameter, $D_{void}$ is obtained for two different energetic material HMX and TATB. It is shown that the criticality of different energetic materials is significantly dependent on their reactive properties.

## Introduction

The meso-scale structure of heterogenous energetic materials (HEMs) is replete with defects such as cracks and voids. These voids collapse when the HEMs are subjected to shock loading, leading to the formation of high temperature localized regions known as the hot-spots. Chemical reaction is initiated in HEMs once the formed hot-spot intensity (temperature and size) exceeds a threshold value. Sensitivity and initiation of a given HEM depends on the dynamics of void-collapse and criticality of the hot spots at the meso-scale. Therefore, it is important to understand and establish the criticality criterion of hot spots.

Tarver et al.[1] proposed critical hot spot curves for HMX and TATB which are based on the hot-spot size and temperature. Tarver et al.[1] assumed a cylindrical/spherical hot-spot with a uniform temperature of a certain size in an otherwise unreacted material (HMX or TATB). The criticality criterion of the hot spots for different combinations of size and temperature was established by solving a reaction-diffusion system. If there is insufficient energy in the hotspot (the hotspot is too small or of low intensity, i.e. temperature) diffusion may transport heat away from the hotspot before complete consumption of the surrounding HEM material. Tarver and coworkers[1] determined the boundary between the go/no-go regions, corresponding to "critical" hotspots. Large hotspots (millimeter sized ones that may arise in impact scenarios) may take much longer to reach criticality. Smaller hotspots that are micron-sized take much shorter times to reach explosion (order of micro-seconds or sub-microseconds). Note that the criticality of the smaller hotspots requires higher temperature.

While the Tarver hotspot curve[1] indicates whether a hotspot is critical or not, it does not indicate how that hotspot was created in the first place. It is important to know what void size must collapse under what loading condition to form a hotspot of critical size and temperature. This is an issue that is addressed in this work by examining the regions in the parameter space – defined by shock pressure $P_s$ and void size $D_{void}$ – where void collapse is sub- and super-critical. This allows for prediction of the criticality hypersurface $\Sigma_{cr} = \Sigma (P_s, D_{void})$. This criticality criterion has the advantage of being defined in terms of operating conditions of a HEM sensitivity experiment, rather than in terms of hotspot size and temperature.

The criticality criterion, $\Sigma_{cr} = \Sigma (P_s, D_{void})$ is dependent on the material constitutive and reactive properties of the energetic material. Using the criticality criteria for HMX and TATB, the relative importance of constitutive and reaction kinetics on criticality condition is also established in the current work.

**Methods**

In the current Eulerian framework, the governing equations are comprised of a set of hyperbolic conservation laws corresponding to the conservation of mass, momentum and energy:

$$\frac{\partial \rho}{\partial t} + \frac{\partial(\rho u_i)}{\partial x_i} = 0 \tag{1}$$

$$\frac{\partial(\rho u_i)}{\partial t} + \frac{\partial(\rho u_i u_j - \sigma_{ij})}{\partial x_j} = 0 \tag{2}$$

$$\frac{\partial(\rho E)}{\partial t} + \frac{\partial(\rho E u_j - \sigma_{ij} u_i)}{\partial x_j} = 0 \tag{3}$$

where $\rho$, and $u_i$, respectively denote the density, and the velocity components, $E = e + 0.5\, u_i u_i$ is the specific total energy, and $e$ is the specific internal energy of the mixture. The Cauchy stress tensor, $\sigma_{ij}$, is decomposed into volumetric and deviatoric components, i.e.,

$$\sigma_{ij} = S_{ij} - p\delta_{ij} \tag{4}$$

The deviatoric stress tensor, $S_{ij}$, is evolved using the following evolution equation:

$$\frac{\partial(\rho S_{ij})}{\partial t} + \frac{\partial(\rho S_{ij} u_k)}{\partial x_k} + \frac{2}{3}\rho G D_{kk}\delta_{ij} - 2\rho G D_{ij} = 0 \tag{5}$$

where, $D_{ij}$ is the strain rate tensor, and $G$ is the shear modulus of material.

Constitutive and Reaction Models for HMX

The material models for HMX used to perform the current analysis are based on the work of Menikoff et al.[2]. A Birch-Murnaghan equation of state is used for the dilatational response of HMX. The equation of state properties are provided in Menikoff et al.[3]. The deviatoric response is obtained by modeling the perfectly plastic behavior of HMX under shock loading. Void collapse under shock loading can lead to the melting of HMX; therefore thermal softening of HMX is modeled using the Kraut-Kennedy relation with model parameters provided in the work of Menikoff et al.[2]. Once the temperature exceeds the melting point of HMX the deviatoric strength terms are set to zero. The specific heat of HMX is known to change significantly with temperature. The variation of specific heat is modeled as a function of temperature as suggested by Menikoff et al.[3]. The chemical decomposition of HMX is modeled using Tarver-Nichols 3 step reaction model[1]. A detailed description of the Eulerian solver and the implementation of the HMX constitutive models is presented in the previous work[4, 5].

Constitutive models for TATB

Equation of state

The hydrostatic pressure, $p$, in Eq. (4) is obtained from the Mie-Grüneisen form of the equation of state:

$$p(\rho, e) = p^s(\rho) + \rho\, \Gamma\, [e - e^s(\rho)] \tag{6}$$

where $\rho$ is the density, $\Gamma$ is the Mie-Grüneisen parameter, and $p^s(\rho)$ and $e^s(\rho)$ are the isentrope pressure and specific internal energy, respectively.

For TATB, the Davis equation of state form is used to find the isentrope pressure[6]:

$$p^s(\rho) = \begin{cases} \hat{p}\left[\exp(4By) - 1\right] & ; \rho < \rho_o \\ \hat{p}\left[\sum_{j=1}^{3}\frac{(4By)^j}{j!} + C\frac{(4By)^4}{4!} + \frac{y^2}{(1-y)^4}\right] & ; \rho \geq \rho_o \end{cases} \quad (7)$$

where $y = 1 - \rho_o/\rho$, $\rho_o$ is the reference ambient density, $\hat{p} = \rho_o A^2/4B$, and A, B, and C are material dependent parameters and their values are obtained from Davis et al.[6]. The isentrope specific internal energy for HMX and TATB is defined as:

$$e^s(\rho) = e_0 - \int_{1/\rho_0}^{1/\rho} p^s(\rho)\, d\left(\frac{1}{\rho}\right) \quad (8)$$

where $e_0$ is the integration constant and represents the stored chemical energy of the explosive.

The temperature in the HMX is obtained from the calculated internal energy in Eq. (8) using the relationship[7]:

$$T(e,\rho) = T^s(\rho)\left(\frac{1+\alpha_{st}}{C_v^0 T^s(\rho)}[e - e^s(\rho)] + 1\right)^{\frac{1}{1+\alpha_{st}}} \quad (9)$$

where,

$$T^s(\rho) = \begin{cases} T_o\left(\frac{\rho}{\rho_o}\right)^{\Gamma^o} & ; \rho < \rho_o \\ T_o \exp(-Zy)\left(\frac{\rho}{\rho_o}\right)^{\Gamma^o} & ; \rho \geq \rho_o \end{cases} \quad (10)$$

The Grüneisen coefficient for the TATB is given by[6, 7],

$$\Gamma(\rho) = \begin{cases} \Gamma^o & ; \rho < \rho_o \\ \Gamma^o + Z y & ; \rho \geq \rho_o \end{cases} \quad (11)$$

where $\Gamma_r^o$ the reference Grüneisen coefficient equals to 0.8168[6], Z is a constant equal to 0.3093[6]. The specific heat for TATB is calculated as[6]:

$$C_v(\rho, T) = C_v^o \left(\frac{T}{T^s(\rho)}\right)^{\alpha_{st}} \quad (12)$$

where $C_v^o$ is the reference specific heat which equals to 837 J/Kg.K[8], $\alpha_{st}$ is a parameter that determines the specific heat change with respect to temperature and its value taken from Davis et al.[6] to be 0.7331.

Strength model for TATB

The TATB material is assumed to exhibit perfectly plastic behavior. The deviatoric part of the stress tensor, $S$ (Eq. (4)) is modeled to capture the perfectly plastic material response. The radial return algorithm for a perfectly plastic material given by Ponthot[9] is used in the current framework. The radial return algorithm evolves the stress deviator, $S$ as an elastic response first using a predictor step (Eq. (5)). Then the stress deviator is mapped back to the yield surface using a corrector step that enforces the plastic flow rule. The values of yield stress, $Y$ and shear modulus, $G$ for TATB used in the calculations are obtained from the work of Najjar et al.[8].

Under high shock compression HEMs can reach melting temperatures and lose their strength. The degradation of TATB strength under melting is modeled in the current framework The melting temperature is taken to be a constant value for the TATB, which is 623 K[8]. Therefore, as the temperature of the HMX reaches the melt temperature the deviatoric component of the stress tensor is set to zero.

Reaction modeling of TATB

The thermal decomposition of TATB are modeled using a multistep chemical kinetic model proposed by Tarver et al.[1]. Chemical decomposition of TATB takes place in 3 steps involving four different species. The three steps are given as,

Reaction1:
TATB ($C_6H_6N_6O_6$) → Solid 2: (13)
furoxans, furazans, ($H_2$, and $H_2O$)

Reaction2:
Solid 2 → Solid 3: furazans, (14)
other fragments, ($H_2O$)

Reaction3:
Solid 3 → final gases ($N_2, CO, CO_2, H_2O$) (15)

The rate equations for all the species in the TATB decompositions are given as,

$$\dot{Y}_1 = -Y_1 Z_1 exp\left(-\frac{E_1}{RT}\right) \quad (16)$$

$$\dot{Y}_2 = Y_1 Z_1 exp\left(-\frac{E_1}{RT}\right) - Y_2 Z_2 exp\left(-\frac{E_2}{RT}\right) \quad (17)$$

$$\dot{Y}_3 = Y_2 Z_2 exp\left(-\frac{E_2}{RT}\right) - Y_3^2 Z_3 exp\left(-\frac{E_3}{RT}\right) \quad (18)$$

$$\dot{Y}_4 = Y_3^2 Z_3 exp\left(-\frac{E_3}{RT}\right) \quad (19)$$

where, $Y_i$ is the mass fraction of the $i^{th}$ species, $Z_i$ is the frequency factor for each reaction, $E_j$ is the activation energy for each reaction, $R$ is the universal gas constant and T is the temperature. The values for each of these constants are obtained from the work of Tarver et al.[1]. The change in temperature because of the chemical decomposition of TATB is calculated by solving the evolution equation,

$$\rho C_p \dot{T} = \dot{Q}_R + \lambda \Delta T \quad (20)$$

where, $\rho$ is the density of TATB, $C_p$ is the specific heat of the reaction mixture, $T$ is the temperature, $\lambda$ is the thermal conductivity of the reaction mixture, $\Delta$ is the Laplacian operator and $\dot{Q}_R$ is the total heat release rate from all the reactions (Equations (13-15) and (16-19)) and given as,

$$\dot{Q}_R = \sum_{I=1}^{3} Q_I \dot{Y}_I \quad (21)$$

where, $I = 1 - 3$ is the reaction number (Equations (16-19)), $Q_I$ is the energy release from each of the reactions and its values are obtained from the work of Tarver et al.[1]. The values of $\lambda$ and $C_P$ for the reaction mixture are obtained by weighted mass fraction average of the specific heat and thermal conductivity for the four species,

$$C_P = \sum_{i=1}^{4} C_P^i Y_i \quad (22)$$

$$\lambda = \sum_{i=1}^{4} \lambda_i Y_i \quad (23)$$

where, $C_P^i$ and $\lambda_i$ are the specific heat capacity and thermal conductivity for the four species obtained from the work of Tarver et al.[1].

**Results**

This work is aimed towards understanding and developing criticality criterion for two different energetic materials HMX and TATB. In the previous work[4], detailed verification and validation of the numerical framework focusing on the void collapse in HMX is presented. In this work, the implementation of the constitutive and reaction models for TATB is verified. The results section is organized as follows, the verification of the numerical framework to perform the void collapse simulation of TATB is presented first, following that the approach to obtain the criticality condition for HMX and TATB is discussed and finally the criticality criterion for HMX and TATB is compared.

Verification of the Numerical Framework

The verification of the numerical framework to perform void collapse analysis in TATB is presented in this section. The results obtained from the void collapse analysis is compared against the simulation results of Najjar et al.[8]. Najjar et al.[8] performed void collapse analysis for a 2 µm diameter single cylindrical void under a sustained shock load of 6 $GPa$ using ALE3D code[10]. Fig. 1 compares the time variation of maximum temperature in the primary jet collapse region[4] obtained from the current work and Najjar et al.[8] simulation. The predictions from the current analysis is in good agreement with the simulations of Najjar et al.[8].

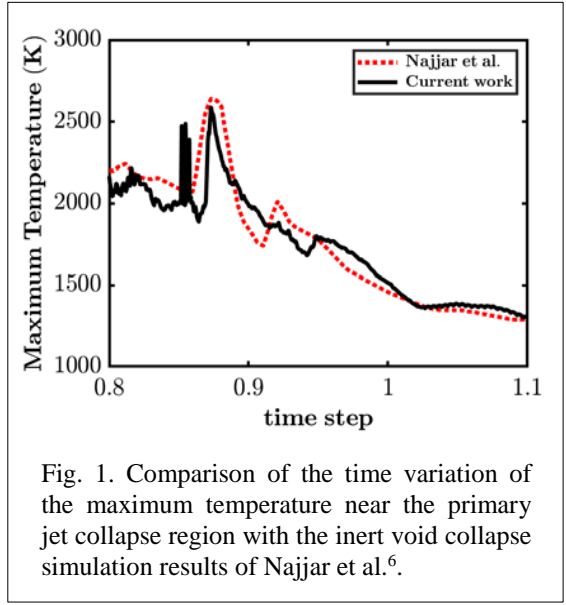

Fig. 1. Comparison of the time variation of the maximum temperature near the primary jet collapse region with the inert void collapse simulation results of Najjar et al.[6].

Having verified the numerical framework for the inert void collapse analysis, the next step is to verify the implementation of the reactive model for TATB decomposition. The 3-step chemical kinetic model of Tarver et al.[1] for TATB decomposition is used in the current analysis. In the work of Tarver et al.[1], the 3-equation reaction kinetics for TATB was used to obtain a criticality curve for TATB initiation. The criticality curve was obtained by performing reaction-diffusion calculations in an otherwise uniform HMX material containing a hot spot. The calculations were performed by instantaneously heating a hot spot of a certain shape and diameter ($D_{hs}$) (cylindrical, spherical or planar) at a specified temperature ($T_{hs}$). The HMX surrounding the hot spot was kept at room temperature (293K). For verification, the current work obtains this critical hot spot curve using the reaction kinetics implementation for cylindrical hot spots. To this end, the reaction diffusion calculations to compute the evolution of a cylindrical hot spot are performed by solving Eq. (20). The criterion for the critical hot spot is obtained by repeating these calculations for different hot spot sizes. The threshold curve is in good agreement with the Tarver et al.[1] result for prediction of the critical condition (Fig. 2). This verifies the parts of the implementation that pertain to chemical kinetics modeling and diffusive transport.

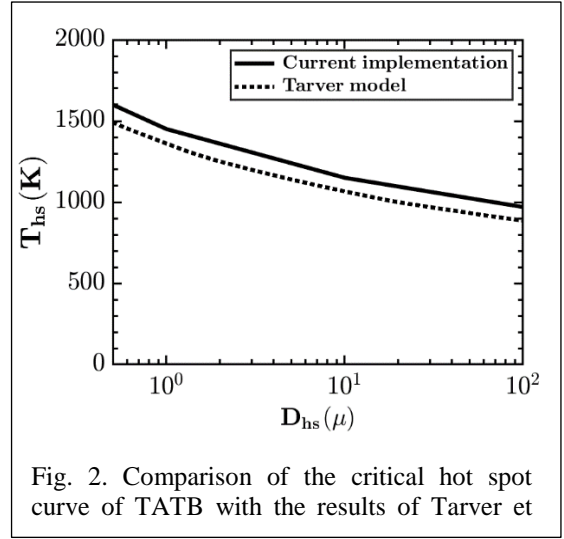

Fig. 2. Comparison of the critical hot spot curve of TATB with the results of Tarver et

Criticality Criterion for HMX and TATB

The aim of the current work is to understand the relative importance of constitutive and reactive properties on the criticality criterion of two different energetic materials, i.e. HMX and TATB. The physics governing the differences in the criticality of the two material is explained in this section.

In the past, Tarver et al.[1] obtained critical hot spot curve for HMX and TATB by performing reaction-diffusion analysis as shown in Fig. 2. Tarver et al.[1] critical hot spot criterion is expressed as a function of hot spot size, $D_{hs}$ and hot spot temperature, $T_{hs}$. Although, their criticality criterion is useful and has been used by modelers[11,12], it involves quantities ($T_{hs}, D_{hs}$) that are not operating parameters in a SDT experiment. It will be more useful to develop a criticality criterion that is a function of controllable input parameters such as shock load, $P_s$ and void diameter, $D_{void}$. Therefore, in this section a criticality criterion for cylindrical voids is derived as a function of shock strength and void size.

Criticality of a hot spot is governed by two competing mechanisms, reaction growth and thermal diffusion. A hot spot will go critical if the reaction time scale $\tau_{reaction}$ is smaller than the thermal diffusion time scale $\tau_{diffusion}$, i.e. depending on the ratio:

$$\psi_1 = \frac{\tau_{reaction}}{\tau_{diffusion}} \qquad (24)$$

Criticality requires $\psi_1 < 1$. Therefore, if an estimate of reaction time scale, $\tau_{reaction}$ and diffusion time scale can be obtained then criticality conditions can be established.

Estimate of time to ignition, $\tau_{reaction}$ for HMX and TATB

The reaction time scale, $\tau_{reaction}$ can be expressed as a function of hot spot temperature, $T_{hs}$ for a given chemical kinetic model of HMX and TATB. Therefore, for different $T_{hs}$, $\tau_{reaction}$ is obtained for the Tarver-Nichols 3-step reactive models of HMX and TATB[1]. For a given $T_{hs}$, time to ignition is recorded when the solid HMX/TATB is completely decomposed to form the final gaseous species. The variation of time to ignition, $\tau_{reaction}$ with $T_{hs}$ for HMX and TATB is shown in Fig. 3. An exponential fit to the data points provides $\tau_{reaction}$ as a function of $T_{hs}$ for HMX and TATB (Eq. (25) and (26)). The time to ignition for a given hot spot temperature is lower for HMX as compared to TATB. Therefore, for the same hot spot temperature TATB will reach ignition at a slower rate as compared to HMX.

HMX:

$$\tau_{reaction}^{HMX} = 2 \times 10^9 e^{-0.035 T_{hs}} \qquad (25)$$

TATB:

$$\tau_{reaction}^{TATB} = 2.6957 \times 10^4 e^{-0.0109 T_{hs}} \qquad (26)$$

Time scale of thermal diffusion, $\tau_{diffusion}$:

The time scale of thermal diffusion is dependent on the hot spot size ($D_{hs}$) and thermal diffusivity ($\alpha_{HMX} = 6.82 \times 10^{-8} \, m^2/s$, $\alpha_{TATB} = 1.01 \times 10^{-7} \, m^2/s$ )[1]:

$$\tau_{diffusion} \approx \frac{D_{hs}^2}{\alpha} \qquad (27)$$

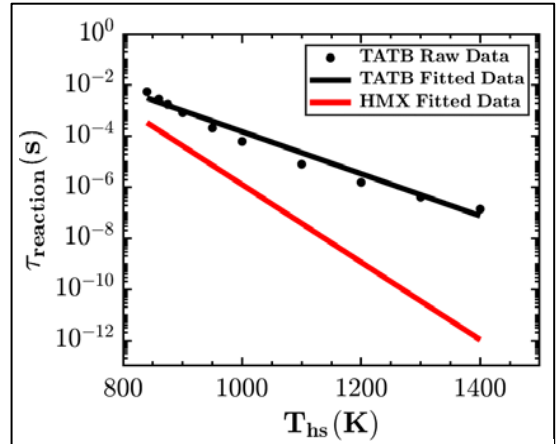

Fig. 3. Comparison of the time to ignition as a function of hot spot temperature for HMX and TATB.

Using Eq. (24) for a critical hot spot:

$$\tau_{reaction} < \tau_{diffusion} \qquad (28)$$

Substituting the expressions for $\tau_{reaction}$ and $\tau_{diffusion}$ from Equations (25) and (27), the criticality condition for Tarver-Nichols model can be obtained as:

$$\tau_{reaction} < \frac{D_{hs}^2}{\alpha} \qquad (29)$$

Eq. (29) provides an expression for the criticality criterion as a function of hot spot characteristics, i.e. $T_{hs}$ and $D_{hs}$. This is similar to the Tarver et al.[1] hot spot criterion. However, as mentioned earlier the aim of the current analysis is to obtain criticality criterion in terms of input parameters that can be controlled. This can be realized if the hot spot characteristics, $T_{hs}$ and $D_{hs}$ can be expressed in terms of input shock pressure, $P_s$ and void size, $D_{void}$. Then Eq. (29) can yield a criticality expression in terms of experimental control parameters.

The hot spot temperature, $T_{hs}$ and size, $D_{hs}$ can be expressed in functional forms as: $T_{hs} = g(P_s, D_{void})$ and $D_{hs} = h(P_s, D_{void})$.

To develop the functional form for $T_{hs}$ and $D_{hs}$, the inert void collapse simulations are performed and analyzed.

$T_{hs}$ as a function of $P_s$ and $D_{void}$ for HMX and TATB

Fig. 4 shows the temperature contours at an intermediate stage of collapse for voids of diameters $D_{void} = 0.5\ \mu m$ and $10\ \mu m$ under two different shock strengths $P_s = 1.27\ GPa$ and $6.4\ GPa$ for HMX. It is clear that $T_{hs}$, for both pressures, i.e. in both plasticity dominated and hydrodynamic collapse modes, is independent of void diameter. The independence of $T_{hs}$ with respect to void diameter is also observed for TATB. This observation simplifies the functional form of $T_{hs}$ to:

$$T_{hs} = g(P_s) \tag{30}$$

To obtain the functional form of $T_{hs}$, the inert void collapse simulation for the $10\ \mu m$ diameter void is performed for input pressures ranging from $2\ GPa - 20\ GPa$. $T_{hs}$ is computed by performing a volume average of temperature field in the hot spot region at the end of the void collapse:

$$T_{hs} = \frac{\int_{V_{hs}} \rho T dV}{\int_{V_{hs}} \rho dV} \tag{31}$$

where $V_{hs}$ is the hot spot volume, $\rho$ is the density and $T$ is the local temperature in the hot spot region. Note that the hot spot in the current work is defined as the region where the temperature is higher than the bulk temperature in the shocked material. However, the minimum value of the hot spot temperature is taken to be 750 K. Fig. 5 shows the simulation data for the variation of the $T_{hs}$ with $P_s$. The figure also shows logarithmic functions fitted to the data. Those functions for HMX and TATB respectively are:

$$T_{hs}^{HMX} = g^{HMX}(P_s)$$
$$= 165.05 * \ln P_s + 718.85 \tag{32a}$$

$$T_{hs}^{TATB} = g^{TATB}(P_s)$$
$$= 196.08 * \ln P_s + 683.9 \tag{32b}$$

where $T_{hs}$ is in K and $P_s$ is in $GPa$.

It is interesting to note that the hot spot temperature values for HMX and TATB are comparable across different pressure ranges. Therefore, the difference in the constitutive properties of HMX and TATB has no significant influence on the hot spot temperature.

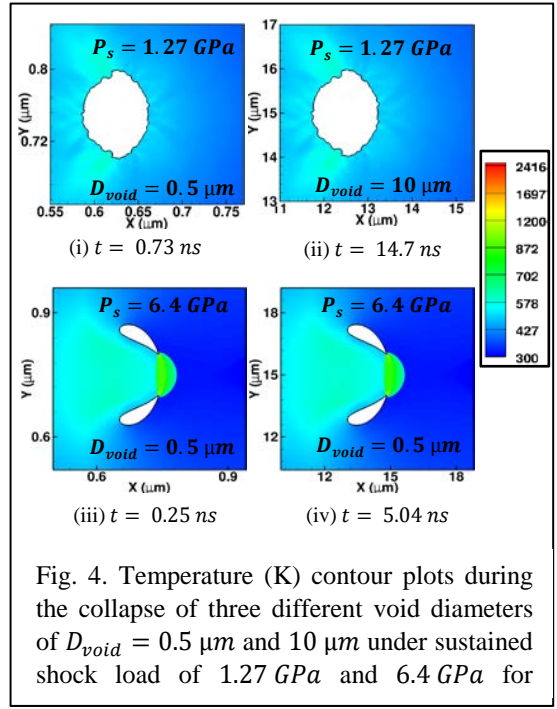

(i) $t = 0.73\ ns$  (ii) $t = 14.7\ ns$

(iii) $t = 0.25\ ns$  (iv) $t = 5.04\ ns$

Fig. 4. Temperature (K) contour plots during the collapse of three different void diameters of $D_{void} = 0.5\ \mu m$ and $10\ \mu m$ under sustained shock load of $1.27\ GPa$ and $6.4\ GPa$ for

$D_{hs}$ as a function of $P_s$ and $D_{void}$

In the above, we show that the hot spot temperature, $T_{hs}$ is independent of void size; the

void collapse modes in Fig. 4 are self-similar for different void sizes under identical loading conditions. To analyze the influence of this self-similar collapse behavior on the hot spot size, the variation of the ratio, $D_{hs}/D_{void}$ with respect to $D_{void}$ is compared for three shock strengths, $P_s = 1.27\ GPa, 2.38\ GPa$ and $6.4\ GPa$ for HMX. The value of $D_{hs} = \sqrt{Area_{hs}}$, with the hot spot defined as mentioned above. The ratio $D_{hs}/D_{void}$ is obtained for the three void diameters viz. 0.5 µm, 10 µm and 30 µm. Fig. 6 shows that $D_{hs}/D_{void}$ varies nonlinearly with respect to shock pressure but linearly with the void diameter. This behavior is also observed for TATB. The functional form for $D_{hs}$ can be expressed as

$$\frac{D_{hs}}{D_{void}} = h(P_s) \qquad (33)$$

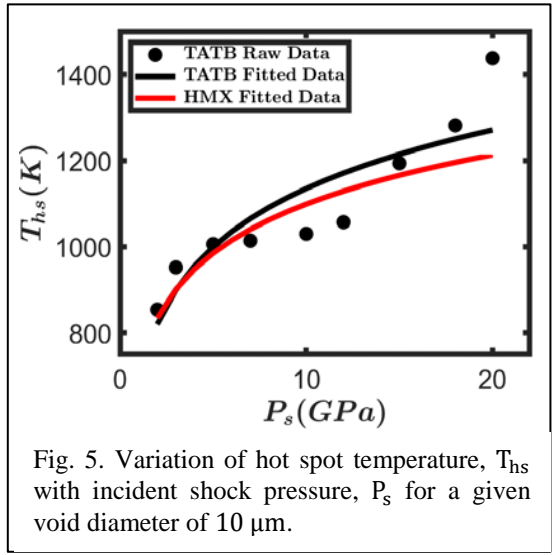

Fig. 5. Variation of hot spot temperature, $T_{hs}$ with incident shock pressure, $P_s$ for a given void diameter of 10 µm.

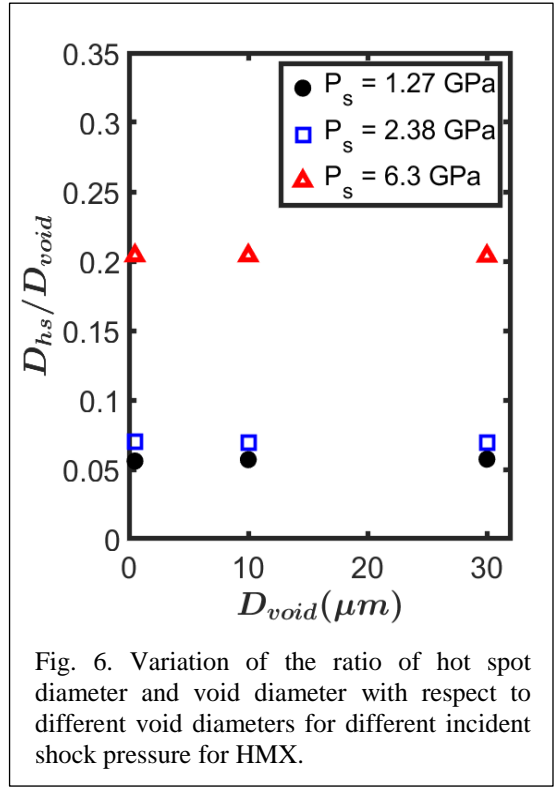

Fig. 6. Variation of the ratio of hot spot diameter and void diameter with respect to different void diameters for different incident shock pressure for HMX.

To obtain $h(P_s)$, the values of $D_{hs}$ for different input shock pressure, $P_s$ are calculated from the inert void collapse simulations for a 10 µm diameter for both HMX and TATB. Fig. 7 shows the variation of $D_{hs}/D_{void}$ with respect to the input shock pressure, $P_s$ for TATB and HMX. The data points are fitted to logarithmic functions here as well to estimate $h(P_s)$ as:

$$\frac{D_{hs}^{HMX}}{D_{void}} = h^{HMX}(P_s)$$
$$= 0.125 * lnP_s - 0.035 \qquad (34a)$$

$$\frac{D_{hs}^{TATB}}{D_{void}} = h^{TATB}(P_s)$$
$$= 0.179 * lnP_s - 0.044 \qquad (34b)$$

where $P_s$ is in $GPa$.

The hot spot diameter for TATB is higher than HMX under similar loading condition. Therefore, TATB forms a larger hot spot as compared to HMX, however the temperature of the hot spot is observed to be in the comparable range.

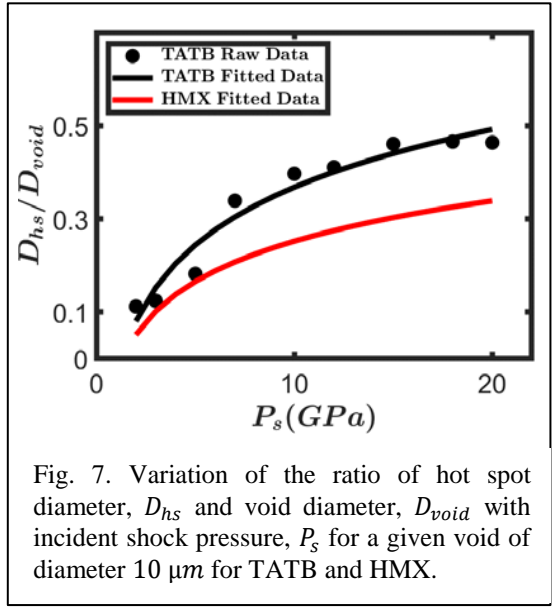

Fig. 7. Variation of the ratio of hot spot diameter, $D_{hs}$ and void diameter, $D_{void}$ with incident shock pressure, $P_s$ for a given void of diameter 10 μm for TATB and HMX.

Criticality as a function of $P_s$ and $D_{void}$ for HMX and TATB

Equations (32) and (34) provide expressions for $T_{hs}$ and $D_{hs}$ as functions of $P_s$ and $D_{void}$. These expressions for $T_{hs}$ and $D_{hs}$ (the (a) part of each equation) are substituted in Equation (29) to yield a hot spot criticality criterion for HMX:

$$2 \times 10^9 e^{-0.035 g(P_s)} < \frac{D_{void}^2 h^2(P_s)}{\alpha_{HMX}} \qquad (35)$$

Rearrangement of Eq. (35) provides a critical void size $D_{critical}$ for HMX as:

$$D_{critical}^{HMX} = \sqrt{\frac{2\alpha_{HMX} e^{-0.035 g^{HMX}(P_s)}}{(h^{HMX})^2(P_s)}} \times 10^9 \qquad (36)$$

Similarly, for TATB (using Equations. (32b) and (34b)) the criticality expression is obtained as,

$$D_{critical}^{TATB} = \sqrt{\frac{2.6957 \alpha_{TATB} e^{-0.019 g^{TATB}(P_s)}}{(h^{TATB})^2(P_s)}} \times 10^4 \qquad (37)$$

Comparison of criticality curve with reactive void collapse simulation data for HMX

Fig. 8 shows the criticality envelope (black curve) in the $(D_{void}, P_s)$ space. A power-law fit can be obtained from the data points and the criticality condition for HMX is expressed as:

$$P_s^{4.45} D_{critical} = 2102 \qquad (38)$$

where, $P_s$ is in $GPa$ and $D_{critical}$ is in μm. The above criticality relationship is obtained from scaling arguments. Using highly resolved reactive meso-scale simulations the criticality criterion as a function of $P_s$ and $D_{void}$ was obtained also for HMX. A detailed description of the reactive Eulerian framework is provided in the previous work[3]. The simulation-based criticality condition is:

$$P_s^{4.38} D_{critical} = 3933 \qquad (39)$$

Fig. 8 compares the simulation-based criticality criterion (Eq. (39), the magenta curve in Fig. 8) and the scaling-based criticality criterion (Eq. (38), the black curve in Fig. 8). The sub-critical and super-critical void collapse data points obtained from the reactive simulations are also shown in Fig. 8.

The scaling-based criticality criterion is in close agreement with the simulation-based criterion. Also, the scaling-based criterion can distinguish the sub-critical and critical regimes with reasonable accuracy.

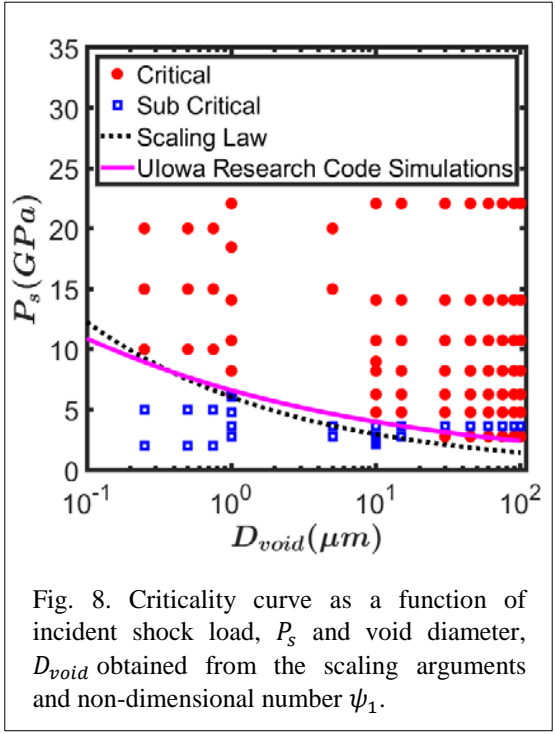

Fig. 8. Criticality curve as a function of incident shock load, $P_s$ and void diameter, $D_{void}$ obtained from the scaling arguments and non-dimensional number $\psi_1$.

Criticality criterion for TATB

Similar to the criticality criterion obtained for HMX (Eq. (38)), using the expressions for time to ignition for the TATB (Eq. (26)), the criticality conditions are obtained. Criticality criterion for HMX and TATB is compared and shown in Fig. 9.

HMX:
$$P_s^{4.45} D_{critical} = 2102 \quad (40)$$

TATB:
$$P_s^{2.38} D_{critical} = 731 \quad (41)$$

Fig. 9 shows the criticality functions in the $P_s - D_{void}$ plane for HMX and TATB. TATB is observed to be insensitive as compared to HMX. In particular, the difference in the sensitivity of TATB and HMX varies greatly for the smaller sized voids.

Conclusions

In the present work, the differences in the criticality prediction of single cylindrical voids for HMX and TATB is analyzed. A numerical framework to perform inert/reactive void collapse simulations of HMX and TATB material is presented. Using a scaling based approach, criticality conditions for both materials are obtained as a function of input shock pressure and void diameter. The scaling based approach is shown to be an effective way to obtain criticality conditions with desirable accuracy. For HMX and TATB, the time to ignition as a function of temperature is obtained. The time to ignition for TATB is observed to be higher as compared to HMX for a given temperature. This is reflected in the criticality predictions of the two materials, where TATB is observed to be less sensitive, although the hot spot size and temperature for both materials are in the comparable range. Therefore, the difference in the constitutive properties of the two material has no significant influence on the criticality. The difference in the criticality of HMX and TATB is attributed to differences in their respective reactive properties.

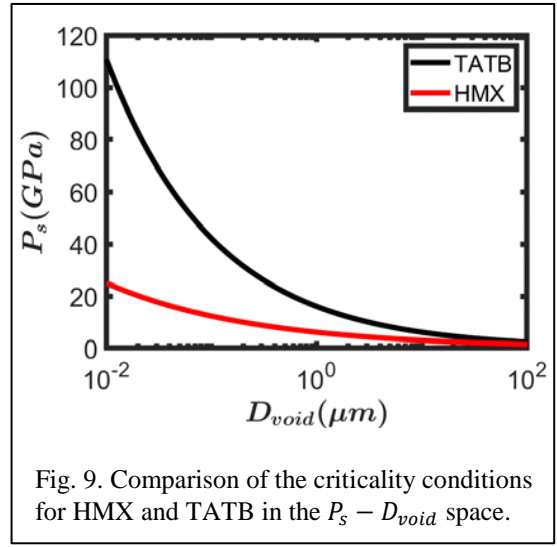

Fig. 9. Comparison of the criticality conditions for HMX and TATB in the $P_s - D_{void}$ space.